\documentclass[twocolumn,prb,showpacs]{revtex4}
\usepackage{graphicx}

\begin{document}
\title{The critical Binder cumulant in a two--dimensional anisotropic
Ising model with competing interactions}
  \author{W.~Selke}
  \affiliation{Institut f\"ur Theoretische Physik,
    RWTH Aachen, and JARA-SIM, 52056 Aachen, Germany}
  \author{L.~N.~Shchur}
  \affiliation{Landau Institute for Theoretical Physics,
    142432, Chernogolovka, and Moscow University of Physics and Technology (MFTI), Dolgoprudny, Russia}

\begin{abstract}

The Binder cumulant at the phase transition of Ising models on
square lattices with ferromagnetic couplings between nearest
neighbors and with competing antiferromagnetic couplings between
next--nearest neighbors, along only one diagonal, is
determined using Monte Carlo
techniques. In the phase diagram a disorder line occurs separating
regions with monotonically decaying and with oscillatory spin--spin
correlations. Findings on the variation of the critical cumulant
with the ratio of the two interaction strengths are compared to
related recent results based on renormalization group calculations.
\end{abstract}


\maketitle

In the field of phase transitions and critical phenomena,
the fourth order cumulant of the order parameter \cite{Binder}, the
Binder cumulant $U$, plays an important role. In particular, the
cumulant may be used to locate the phase transition
from the intersection of the cumulant for different system
sizes. The cumulant also allows to compute the critical exponent of the
correlation length, and thence to identify the universality class
of the transition.

However, care is needed when attempting to identify the universality
class from the value of the Binder cumulant, in the thermodynamic
limit, at the transition, $U^*$. Indeed, that value is known to
depend, in a given universality class, on various aspects including
the boundary conditions, the shape of the system (being extrapolated
to the thermodynamic limit), and the anisotropy of the correlations
or interactions. \cite{Binder,Bloete,Janke,CD,SelSh,Selke,Dohm} On
the other hand, $U^*$  may not depend on other details of the
system like the spin value \cite{Bruce} or the
lattice structure. \cite{Selke}

In recent years, based on renormalization group calculations
of Dohm and Chen \cite{CD,Dohm} and subsequent Monte Carlo
simulations \cite{SchDr,SelSh,Selke,SP}, the influence
of anisotropic interactions on the critical Binder
cumulant $U^*$ has been
elucidated. Much attention has been focused on the
Ising model with nearest neighbor (nn) couplings, $J$, where
the anisotropy is introduced
by the next--nearest (nnn) couplings along only one
diagonal of the lattice, $J_d$ \cite{CD,SelSh,Dohm}. One
encounters a 'nondiagonal
anisotropy matrix' \cite{CD}. Then, in general, there
is no simple transformation relating  $U^*$, for given
boundary condition and shape, to that
of the isotropic model by adjusting the shape \cite{CD}. Such
a transcription may be easily performed in the
case of a diagonal
anisotropy matrix, as it occurs, for instance, for
the nn Ising on a square lattice with different
vertical and horizontal ferromagnetic interactions, where
$U^*$ of the anisotropic model on lattices with
square shape may be expressed by $U^*$ of the isotropic
model on lattices with rectangular shapes. \cite{Bloete,SelSh}

In this contribution, we shall extend our previous Monte Carlo study
for anisotropic nn and nnn Ising models on square lattices with only
ferromagnetic interactions, $J, J_d >0$, to the case of competing
nnn antiferromagnetic couplings, $J_d < 0$, with $J$ remaining
ferromagnetic. Thereby, spatially modulated, oscillatory spin--spin
correlations may occur, adding an interesting feature to the phase
diagram. The present study has been partly motivated by related
recent quantitative renormalization group calculations of
$U^*(J_d/J)$ for a closely related model, showing intriguing symmetry
properties \cite{Dohm}, as will be discussed below.

The Hamiltonian of the model may be written in the form

 \begin{equation}
{\cal H} = -\sum\limits_{x,y} S_{x,y} ( J (S_{x+1,y}
 + S_{x,y+1}) + J_d S_{x+1,y+1}))
\end{equation}

\noindent
where $S_{x,y}= \pm 1$ is the Ising spin at site $(x,y)$. $J>0$ is
the ferromagnetic nn coupling along the principal
axes of the square lattice, while the nnn coupling, $J_d$ acts
along only one diagonal, i.e. along
the [11] direction of the lattice. $J_d$ may be
ferromagnetic, $J_d >0$, as has been studied before \cite{SelSh},
or antiferromagnetic, $J_d <0$.

Before turning to the critical Binder cumulant, we shall first
discuss the phase diagram, exhibiting
interesting features, especially for antiferromagnetic
nnn couplings.

At $J_d/J> -1$, there is
a ferromagnetic phase at low temperatures. The exact transition
temperatures, $k_BT_c/J$, are known to be determined
by \cite{Berker,Houtappel}
\begin{equation}
 (\sinh(2J/k_BT_c))^2 +2\sinh(2J/k_BT_c)\sinh(2J_d/k_BT_c) = 1,
\end{equation}
\noindent
The line is depicted in Fig. 1. The  transition temperature $T_c$
goes to zero on approach to $J_d/J =-1$. At that point, the
ground state is highly
degenerate, with the energy per site being $E_0$= $-J$. The
degenerate configurations include the ferromagnetic
structures, uncoupled antiferromagnetic Ising
chains along the [11] direction, and horizontal and vertical
ferromagnetic stripes of spins with alternating sign corresponding
to coupled modulated Ising chains along the [11] direction. A
rather large degeneracy, due to
the uncoupled antiferromagnetic chains along the [11] direction, persists at
$J_d/J < -1$. Indeed, it has been argued that there is
no long--range ordering at low temperatures in that part of the
phase diagram \cite{Wu}. Actually, in the following we shall
consider $J_d/J \ge -1$.
 \begin{figure}[h]\centering
        \includegraphics[angle=0,width=\columnwidth]{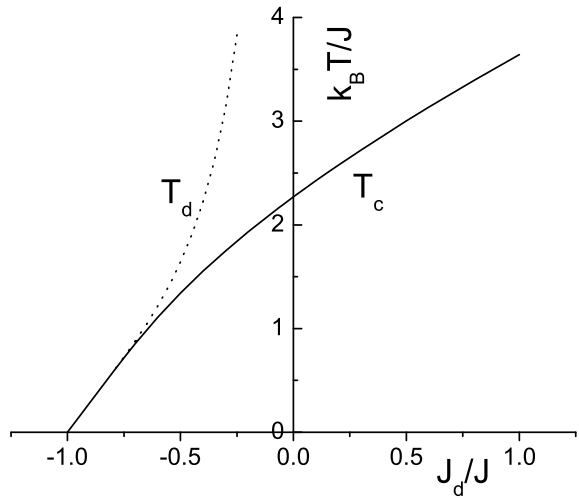}
         \caption{\label{fig1} Phase diagram of Hamiltonian (1), showing
  the exact boundary line of the ferromagnetic phase \cite{Houtappel} (solid
  line) and the disorder line \cite{Stephenson} (dashed line).}
\end{figure}

Due to the competing ferro-- and antiferromagnetic
couplings, at $J_d/J <0$, a disorder line, $T_d(J_d/J)$, separates,
above $T_c$, the region with only monotonically decaying
spin--spin correlations from the one with oscillatory
correlations \cite{Stephenson}. The disorder line of the model, eq. (1), has
been calculated exactly as well. \cite{Stephenson,Peschel} It is
determined by
\begin{equation}
 \cosh(2J/k_BT_d)= \exp(-2J_d/k_BT_d).
\end{equation}
\noindent The line $T_d(J_d/J)$ is also shown in Fig. 1. It arises
from the highly degenerate point at ($J_d/J=-1, T=0)$, reflecting
the spatially modulated configurations occurring there as ground
states. At low temperatures, $-J_d$ being not far from $J$, the disorder line
$T_d$ follows closely the phase transition line, $T_c$, eventually
moving away from it towards higher temperatures, as $J_d$ gets
weaker, and finally approaching infinite temperature at vanishing
nnn antiferromagnetic couplings, see Fig. 1. Of course, there is no disorder
line for ferromagnetic nnn interactions, $J_d>0$.

The different types of correlations at temperatures
below and above the disorder line, fixing $J_d/J$, are
illustrated in Fig. 2, displaying  Monte Carlo data for
spin--spin correlations along the [11]
direction, G$_1$(r)= $<S_{x,y}S_{x+r,y+r}>$, along the
principal axes, G$_2$(r)= $<S_{x,y}S_{x,y+r}>= <S_{x,y}S_{x+r,y}>$, and
perpendicular to the [11] direction, G$_3$(r)=
$<S_{x,y}S_{x+r,y-r}>$. At $T> T_d$, G$_1$
and G$_2$ decay, for sufficiently large distances $r$,
exponentially and in an oscillatory, purely sinusoidal
manner, with the wavenumber depending on $J_d/J$ and
temperature \cite{Stephenson}. At and below the
disorder line, the correlations along the principal axes and
along the [11] direction
decay monotonically. In addition, our
simulations suggest that G$_3$ decays also above
$T_d$ monotonically (and exponentially) with
distance, indicating that the competing interactions do not
affect qualitatively correlations perpendicular to the
[11] direction.

\begin{figure}[h]\centering
        \includegraphics[angle=0,width=\columnwidth]{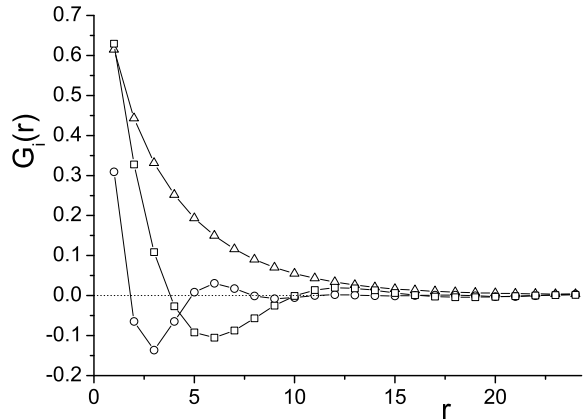}
\caption{\label{fig2}Simulated spin--spin correlation function G$_i$, with
 $i$= 1 (circles), 2 (squares), 3 triangles), as described
 in the text, at $J_d/J= -3/4$ and $k_BT/J= 0.9$, for lattices
 with $48^2$ spins.}
\end{figure}

Note that disorder lines also exist in other
Ising models with competing interactions, for instance, in the
much studied ANNNI model. \cite{Review,Becc}

Let us now turn to the Monte Carlo findings on the critical Binder
cumulant $U^*$ of the model, eq. (1). $U^*$ is defined by \cite{Binder}

\begin{equation}
  U^*= U(T_c) = 1- <M^4>/(3 <M^2>^2)
\end{equation}

\noindent
taking the thermodynamic limit;
$<M^2>$ and $<M^4>$ denote the second and fourth moments
of the order parameter, the magnetization $M$.

To estimate $U^*(J_d/J)$, we simulated the model
for square shapes with
$L^2$ sites or spins, employing full periodic boundary
conditions, using the standard Metropolis algorithm (note
that, e.g., cluster flip algorithms are usually rather inefficient in
case of competing ferro-- and antiferromagnetic
interactions). Monte Carlo runs with, typically,
$5\times 10^8$ Monte Carlo steps per site were
performed, averaging then over several, up to ten, of these
runs to obtain final estimates and to determine statistical
error bars. $L$ ranged from 4 to 64.  To extrapolate to the
thermodynamic limit, $L \longrightarrow \infty$, least square
fits were done. The procedure is
exemplified in Fig. 3. The final error bars result from
the fits (one may emphasize that the finite--size behavior of
the Binder cumulant is not known
in the anisotropic case).

\begin{figure}
  \begin{center}
        \includegraphics[angle=0,width=\columnwidth]{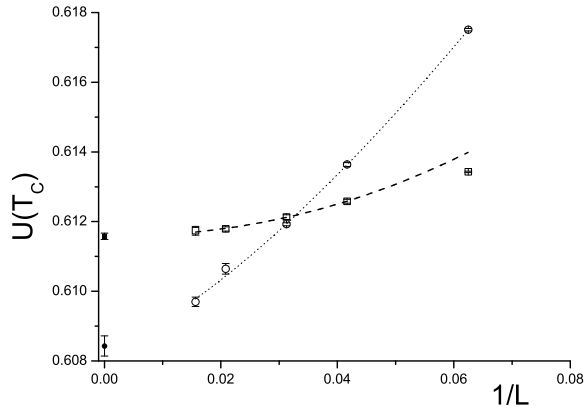}
  \end{center}
  \caption{\label{fig3} Binder cumulant at $T_c$, $U(T_c)$,  versus
   inverse system size $1/L$, including least square fits to
   the Monte Carlo data (dashed lines) and the resulting estimates
   for the critical Binder cumulant, $U^*$. The cases $J_d/J= -0.316$
   (squares) and $-5/9$ (circles) are shown.}
\end{figure}

In Fig. 4, we plot $U^*$ against $s= 1/(1+(J/J_d))$, following a
recent renormalization group analysis \cite{Dohm}, with $s < 0$ for
$J_d < 0$ and $s > 0$ for $J_d > 0$. The reason for this choice will
be discussed below. We also include our previous estimates
\cite{SelSh} of $U^*$ for $J_d \ge 0$, adding one additional new
point at $s= 1/6$. Obviously, the critical Binder cumulant depends,
both in the case of only ferromagnetic interactions and also in the
case of competing antiferromagnetic couplings, continuously on
$s$ (or $J_d/J$). The
transition at $T_c(s)$ is always in the Ising universality class, as
we confirmed by monitoring, especially, the, asymptotically,
logarithmic size dependence of the specific heat at $T_c$. Thence,
$U^*$, for given boundary condition and shape, changes, within the
Ising universality class, with the anisotropy, $J_d/J$.

Interestingly, $U^*(s)$ is (almost) symmetric around the
isotropic situation, $s=0$, at small values of $|s|$. However, for
larger values of $|s|$, there is a pronounced asymmetry. This
may be understood by the observation  that  the model reaches
one--dimensional limits, where $U^* \longrightarrow 0$, for
$s=1$ or $J_d/J= \infty$ at positive $s$, and
for $s= -\infty$ or $J_d/J= -1$ at negative $s$. The nonmonotonicity of $U^*$
for fairly small values of $|s|$ is also worth mentioning. It has
been first found for positive $s$ in a previous
Monte Carlo study \cite{SelSh}, and it also holds for negative $s$.

\begin{figure}
  \begin{center}
        \includegraphics[angle=0,width=\columnwidth]{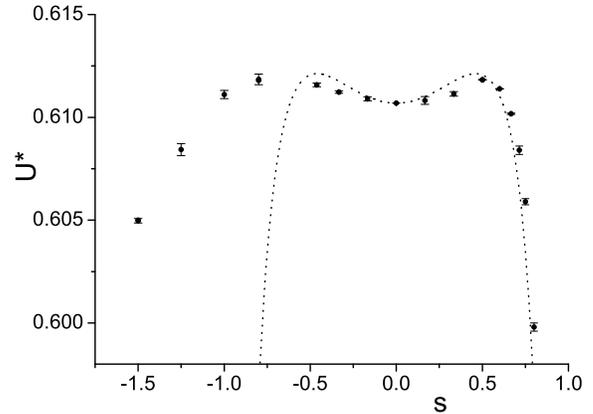}
  \end{center}
  \caption{\label{fig4} Critical Binder cumulant, $U^*(s)$, versus
   $s= 1/(1+(J/J_d))$. The result of a renormalization group analysis
   \cite{Dohm}, see text, is included as dashed line.}
\end{figure}

The previous Monte Carlo data \cite{SelSh} on $U^*(s)$ at $s \ge 0$
agree closely with results of a recent renormalization group (RNG)
analysis by Dohm, as has been already discussed there \cite{Dohm}.
That analysis has been done for the model, eq. (1), augmented by a
ferromagnetic interlayer nn coupling $J_l$ on a cubic lattice, with
$J_l= J+J_d$. The interlayer coupling has been chosen to preserve
essential features of the anisotropy matrix of the
model on the square lattice. \cite{Dohm} Indeed, this choice allows to
describe well, in the framework of RNG theory, the
simulational data on $U^*(s)-U^*(0)$ for the two--dimensional model
with ferromagnetic nnn interactions, $s \ge 0$. \cite{Dohm}
Certainly, the critical Binder cumulant in the isotropic case,
$s=0$, is different for square and cubic lattices, thereby
motivating to look at the difference, $U^*(s)-U^*(0)$, of the
critical cumulants \cite{Dohm}. The RNG study yields a perfect
symmetry of $U^*(s)-U^*(0)$ around $s=0$, and an interesting
nonmonotonic behavior in $s$.

To compare the present Monte Carlo
data to the RNG findings, we prefer to
plot, see Fig. 4, the RNG results in
the form $U^*_{RNG}(s)U^*(s=0)/U^*_{RNG}(s=0)$, with the critical
Binder cumulant of the two--dimensional isotropic Ising model, $U^*(0)$,
being \cite{Bloete} 0.61069... Of course, the perfect
symmetry and nonmonotonicity found in the RNG study is preserved
by this choice, as displayed in Fig. 4. As follows from that
figure, the symmetry, suggested by the RNG analysis, is approximated
closely in the two--dimensional Ising model with antiferromagnetic
nnn interactions for small
values of $|s|$, confirming the
nonmonotonicity also for negative $s$. But there are
pronounced deviations at stronger
antiferromagnetic nnn couplings.

In fact, for competing antiferromagnetic couplings the phase
diagrams of the models in two and three dimensions, and thence the
corresponding critical Binder cumulants, may be expected to differ
substantially. In particular, in two dimensions there is, as
discussed above, a phase transition of Ising type down to $J_d/J= -1$, with
spatially modulated correlations occurring only in the disordered
phase above the disorder line. In contrast, in three dimensions, the
transition line of Ising type between the ferro- and paramagnetic
phases may be expected to extend only up to a Lifshitz point
\cite{Hornreich,Selke2,Diehl} at $(J_d/J)_{LP}$ ($>-1$) or $s_{LP}$,
at which the ferromagnetic, the spatially long--range ordered
modulated, and the disordered phases meet, similar to the
well--known situation in the ANNNI model. \cite{Review}

Indeed, a standard mean--field calculation shows that the Lifshitz point
occurs, for $J_l= J+J_d$, at $J_d/J= -1/2$, corresponding to
$s_{LP}=-1$. In the modulated phase close to the transition to the
paramagnetic phase, at $J_d/J < (J_d/J)_{LP}= -1/2$, the magnetization along the
$x$-- and $y$-- axes change in a sinusoidal manner. The modulated,
long--range ordered phase seems to arise from the highly degenerate
ground states at $J_d/J= -1$, with the transition line between the
ferromagnetic and the modulated phases, at $-1 < s <-1/2$, being of
first order. The vicinity of that degenerate point may be explored
by systematic low--temperature series
expansions \cite{Fisher}, being however, well beyond the scope of
the present study. Because there are modulated magnetization pattern
along two principal axes of the cubic lattice, the $x$-- and the
$y$--axes, we are dealing here with a 'biaxial Lifshitz point'
($m=2$ in the standard notation \cite{Hornreich}).

The existence of a Lifshitz point at $s=-1$, has been argued to allow
for the complete symmetry of
$U^*(s)$ around $s=$ in the RNG  analysis \cite{Dohm}. Going
beyond mean--field theory, the
location of the biaxial Lifshitz point may shift somewhat. To
determine accurately the position of the Lifshitz
point, high temperature series may be very helpful, as has
been found, for instance, in
the case of the ANNNI and
related models. \cite{Stanley,Oitmaa,Butera} Obviously, it
is an open problem, in which way the possible shift in
the Lifshitz point may affect the proposed perfect symmetry
of $U^*(s)$. Note that the
lower critical dimension of a biaxial Lifshitz point is
three \cite{Hornreich,Shpot}, excluding its existence at
non--zero temperatures for the square lattice, but not
for the cubic lattice.

Certainly, it is desirable to determine $U^*(s)$ of the
three--dimensional model in simulations. However, neither $T_c(s)$ nor
the location of the Lifshitz point are known. To get then
data on $U^*$ of the required high accuracy, a huge amount of computer
time would be needed, being hardly feasible at present.

In summary, the critical Binder cumulant $U^*$ of an anisotropic
two--dimensional Ising model with competing ferro-- and
antiferromagnetic interactions has been determined. Because the
transition is known exactly, one arrives at accurate estimates
based on extensive Monte Carlo simulations. Employing full periodic
boundary conditions and considering square shapes, $U^*$ is found to
vary continuously with changing anisotropy $s$. A remarkably close
agreement of our findings on $U^*(s)$, at positive values of s
and, in case of competing interactiosn, for small negative
values of $s$ with the results of a recent
renormalization group analysis is observed. Differences at stronger
competing anisotropy are explained by the absence of a biaxial
Lifshitz point at non--zero temperatures in two dimensions.\\

\acknowledgments We thank V. Dohm for inspiring discussions as well
as H. W. Diehl and S. E. Korshunov for useful correspondence and
conversation. W.S. thanks the Landau Institute in Chernogolovka for
the very kind hospitality during his sabbatical visit there. L. N.
S. acknowledges partial support by the Program of the Russian
Academy of Sciences and by the grant from Russian Foundation for
Basic Research.


\begin{thebibliography}{99}

\bibitem{Binder} K.\ Binder, Z. Physik\ B\ \textbf{43}, 119 (1981);
  Phys.\ Rev.\ Lett.\ \textbf{47}, 693 (1981).
\bibitem{Bloete} G.\ Kamieniarz and H.\ W.\ J.\ Bl\"ote, J.\ Phys.\ A\ :
  Math.\ Gen.\ \textbf{26}, 201 (1993).
\bibitem{Janke} W.\ Janke, M.\ Katoot, and R.\ Villanova, Phys.\ Rev.\ B \textbf{49}, 9644 (1994)
\bibitem{CD} X.\ S.\ Chen and V.\ Dohm, Phys.\ Rev.\ E\ \textbf{70}, 056136 (2004).
\bibitem{SelSh} W.\ Selke and L.\ N.\ Shchur, J.\ Phys.\ A: Math.\
  Gen.\ \textbf{38}, L739 (2005).
\bibitem{Selke} W.\ Selke, Eur.\ Phys.\ J.\ B\ \textbf{51}, 223 (2006); J.\ Stat.\ Mech.\ P04008 (2007).
\bibitem{Dohm} V.\ Dohm, Phys.\ Rev.\ E\ \textbf{77}, 061128 (2008).
\bibitem{Bruce} D.\ Nicolaides and A.\ D.\ Bruce,  J.\ Phys.\ A: Math.\
  Gen.\ \textbf{21}, 233 (1988).
\bibitem{SchDr} M.\ Schulte and C.\ Drope, Intern.\ J.\ Mod.\ Phys.\ C\ \textbf{16}, 1217 (2005).
\bibitem {SP} M.\ A.\ Sumour, D.\ Stauffer, M.\ M.\ Shabat, and A.\ H.\
  El-Astal, Physica\ A\ \textbf{368}, 96 (2006).
\bibitem{Berker} A.\ N.\ Berker and K.\ Hui, Phys.\ Rev.\ B
  \textbf{48}, 12393 (1993).
\bibitem{Houtappel} R.\ M.\ F.\ Houtappel, Physica\ \textbf{16}, 425
(1950).
\bibitem{Wu} K.\ Y.\ Lin and F.\ Y.\ Wu, Z.\ Phys.\ B
 \textbf{33}, 181 (1979)
\bibitem{Stephenson} J.\ Stephenson, Phys.\ Rev.\ B
  \textbf{1}, 4405 (1970); J.\ Math.\ Phys.\ \textbf{11}, 413 (1970).
\bibitem{Peschel} I.\ Peschel and V.\ J.\ Emery, Z.\ Phys.\ B
 \textbf{43}, 241 (1981).
\bibitem{Review} W.\ Selke, Phys.\ Rep.\ \textbf{170}, 213 (1988).
\bibitem{Becc} M.\ Beccaria, M.\ Campostrini, and A.\ Feo, Phys.\ Rev.\ B\ \textbf{73}, 052402 (2006).
\bibitem{Hornreich} R.\ M.\ Hornreich, M.\ Luban, and S.\ Shtrikman,
  Phys.\ Rev.\ Lett. \textbf{35}, 1678 (1975).
\bibitem{Selke2} W.\ Selke in 'Phase
  Transitions and Critical Phenomena', ed. by C Domb and J L Lebowitz
  (New York: Academic), vol 15, p.1 (1992)
\bibitem{Diehl} H.\ W.\ Diehl, Pramana-J.\ Phys.\ \textbf{64}, 803 (2005).
\bibitem{Fisher} M.\ E.\ Fisher and W.\ Selke, Phys.\ Rev.\ Lett. \textbf{44}, 1502 (1980).
\bibitem{Stanley}  S.\ Redner and H.\ E.\ Stanley, Phys.\ Rev.\ E\ \textbf{16}, 4901 (1977).
\bibitem{Oitmaa} J.\ Oitmaa, J.\ Phys.\ A\ : Math.\ Gen.\ \textbf{18}, 365 (1985).
\bibitem{Shpot}  H.\ W.\ Diehl and M.\ Shpot, Phys.\ Rev.\ B\ \textbf{62}, 12338 (2000).
\bibitem{Butera}  P.\ Butera and M.\ Pernici, Phys.\ Rev.\ B\ \textbf{78}, 054405 (2008).
\end{thebibliography}
\end{document}